\documentclass{article}
\pdfoutput=1 % if your are submitting a pdflatex (i.e. if you have
\usepackage{jcappub,aasmacros} % for details on the use of the package, please
\usepackage[T1]{fontenc} % if needed
\usepackage{graphicx}% Include figure files
\usepackage{dcolumn}% Align table columns on decimal point
\usepackage{bm}% bold math
\usepackage[dvipsnames]{xcolor}
\usepackage{times}
\usepackage{multirow}
\usepackage{mathtools}
\usepackage{threeparttable}
\usepackage{tabularx}
\usepackage{enumitem}

\newcommand{\footnoteref}[1]{\textsuperscript{\ref{#1}}}
\newcommand{\lvb}{\beta}

\title{A Bayesian interpretation of inconsistency measures in cosmology}

\author[a,b]{Weikang Lin}
\author[c]{and Mustapha Ishak}

\affiliation[a]{Tsung-Dao Lee Institute, Shanghai Jiao Tong University, Shanghai 200240, China}
\affiliation[b]{Department of Physics, North Carolina State University, Raleigh, NC 27695, USA}
\affiliation[c]{Department of Physics, The University of Texas at Dallas, Richardson, Texas 75080, USA}

\emailAdd{weikanglin@sjtu.edu.cn}
\emailAdd{mishak@utdallas.edu}

\abstract{Measures of inconsistency and tension between datasets have become an essential part of cosmological analyses. It is important to accurately evaluate the significance of such tensions when present. We propose here a Bayesian interpretation of inconsistency measures that can extract information about physical inconsistencies in the presence of data scatter. This new framework is based on the conditional probability distribution of the level of physical inconsistency given the obtained value of the measure. We use the index of inconsistency as a case study to illustrate the new interpretation framework, but this can be generalized to other metrics. Importantly, there are two aspects in the quantification of inconsistency that behave differently as the number of model parameters increases. The first is the probability for the level of physical inconsistency to reach a threshold which drops with the increase of the number of parameters under consideration. The second is the actual level of physical inconsistency which remains rather insensitive to such an increase in parameters. The difference between these two aspects is often overlooked, which leads to a long-standing ambiguity: when a given  inconsistency is found between two constraints, its ``significance'' seems to be lower when considered in a higher-dimensional parameter space. This ambiguity is resolved by the Bayesian interpretation we introduce in this work because the conditional probability distribution includes all the statistical information of the level of physical inconsistency. Finally, we apply the Bayesian interpretation to examine the (in)consistency between Planck versus the Cepheid-based local measurement, the Dark Energy Survey (DES), the Atacama Cosmology Telescope (ACT) and WMAP. We confirm and revisit the degrees of previous physical inconsistencies and show the stability of the new interpretation with respect to the number of cosmological parameters compared to the commonly used n-$\sigma$ interpretation when applied to cosmological tensions in multi-parameter spaces.  
}

\begin{document}
\maketitle
\flushbottom

\section{Introduction}\label{section-introduction}
The standard cosmological model has been very successful in explaining and making predictions about the observed universe. Joint cosmological analyses of multiple complementary datasets have allowed one to break degeneracies between cosmological parameters and to significantly increase the precision of cosmological constraints. However, Bayesian joint analyses do not automatically detect or show inconsistencies between datasets when present. This needs to be dealt with  since inconsistencies between cosmological constraints have been reported in the literature, including some that have persisted for almost a decade now; see for example Refs.\,\cite{2021-Riess-etal-H0,KiDS-VIKING-DES-2019,Lusso-etal-2019,Wong-etal-2019-H0licow}. 

Debates have been ongoing about the degree of these inconsistencies and what may be their possible causes; see for example the following partial list of references  \cite{2020-Lin-Mack-Hou,Lombriser-2019-H0,Blinov-etal-2019-interacting-v-H0,2018-Gomez-Valent-Amendola,2018Yu-Ratra-Wang,2017-Chen-Kumar-Ratra,WL2017b,2017-Sola-Gomez-Valent-deCruz,2017-Yang-Pan-Mota,2017-Barenhoim-Kinney-Morse,2020-Yao-Shan-Zhang-Kneib,2018-Haridasu-Lukovi-Vittorio,2018-Bolejko.K,2016-Hubble-reconcile,2016-Valentino-Melchiorri-Silk-reconciling-Planck-local,2017Ooba-Ratra-Sugiyama,2018-Park-Ratra,2018-Gomez-Valent-Sola,2018-Kazantzidis-Perivolaropoulos,2018-Park-Ratra2,Vagnozzi-2019,Collett-etal-2019-strong-lesning-SN,Kazantzidis-Perivolaropoulos-2019,Yang-eta-2019-many-w}. Inconsistencies between cosmological observations could signal unaccounted for systematic errors and/or the breakdown of the underlying model or theory. Moreover, using constraints obtained from combining inconsistent datasets puts into question current and future results derived from them. Therefore, it is essential to resolve any inconsistencies between datasets. For a better understanding of the reported inconsistencies and their underlying causes, it has become an important and timely task to study how to properly describe them and quantify their significance. Indeed, this topic has recently attracted a lot of attention and many inconsistency measures have been proposed in the literature; see for example \cite{2003-Maltonia-Schwetz,2006-Marshall-etal-Bayesian,2011Robustness-March-etal,2013-tension-Verde-etal,2015-discordance-MacCrann-etal,2015-Battye-Charnock-Moss-difference-vector,2017-Charnock-Battye-Moss,2014-rel-entropy-Seehars-etal,2016-quantify-concor-Seehars-etal,2016-Grandis-information-Gains,2017MNRAS-Joudaki-etal-DIC-inconsistency,WL2017a,2018-Adhikari-Huterer,2018-Raveri-Hu,Handley-etal-2019-suspiciousness,2020-Handley-Lemos,Park-Rozo-2019,Nicola-Amara-Refregier-2019,Kohlinger-etal-2019}. However,  results and conclusions based on different measures are generally different, even in the simple Gaussian cases\footnote{Gaussian cases here refer to Gaussian functions of model parameters.}.

The inconsistency between cosmological parameter results obtained from different observations can be due to systematic errors, an incorrect underlying model, or data scatters. While the first two have certain physical meanings, the third is purely statistical. We call inconsistency due to systematic errors or an incorrect/incomplete underlying model as physical inconsistencies and those due to data scatter as a statistical inconsistency.  A fundamental difference between physical inconsistencies and statistical inconsistencies is that physical inconsistencies manifest as the same parameter difference in each data realization, but statistical inconsistencies due to data scatter are random and different for each realization drawn from data likelihoods. It is essential to properly interpret the value of an inconsistency measure in the presence of data scatter. 

The way to deal with the data scatter aspect has been an evolving topic. Due to data scatter, measures of inconsistency generally follow some distribution. Earlier, Refs.\,\cite{2015-Battye-Charnock-Moss-difference-vector,2017-Charnock-Battye-Moss} proposed to calculate the significance of the inconsistency measure as the difference between the obtained value of the measure and its mean value normalized by the standard deviation of the measure\footnote{That is,   ${\rm{significance}}=\frac{{\rm{measure}}-<\rm{measure}>}{\sigma_{\rm{measure}}}$.}. Now, a more commonly adopted way in the literature is to calculate the probability to exceed (PTE) and the associated $n$-$\sigma$ value of a given inconsistency estimator that quantifies the probability of that estimator being equal to or larger than the obtained value \cite{2018-Raveri-Hu,Handley-etal-2019-suspiciousness,2018-Adhikari-Huterer}. However, these considerations are not Bayesian as they are based on the probability distribution of an estimator in the absence of any physical inconsistency, i.e., $P(\rm{estimator}|{\rm{physical~ inconsistency}}=0)$. More importantly, as we will show, there is an ambiguity in the quantification of inconsistencies in multiple-parameter models: when a certain inconsistency is found between two constraints, its ``significance'' seems to be lower when considered in a higher parameter space. Such an ambiguity leads to the different assessments of the significance level (the $n$-$\sigma$), e.g., in the $\sigma_8$ tension between Planck and KiDS-1000 when considering different parameter numbers \cite{2020-Heymans-etal}. This ambiguity cannot be resolved in those non-Bayesian interpretations of inconsistency estimators. We will show that the commonly adopted $n$-$\sigma$ notation of significance level underestimates the physical inconsistency in general when there is more than one model parameter. The complication of the quantification in multiple-parameter models have been partially recognized in the literature \cite{WL2017a,2018-Raveri-Hu,2020-Handley-Lemos}, but the essence of the problem was not identified and so far no commonly-accepted resolution was found.\footnote{
Moreover, it is worth noting some important points for inconsistency measures that were pointed out in our early work \cite{WL2017a} and have been re-emphasized in some other recent works like  \cite{2020-Handley-Lemos}. These include for example (1) inconsistency can be hidden when constraints are viewed and compared in marginalized plots; (2) the undesirable dependence on priors of the Bayes evidence ratio in quantification of inconsistencies.}

In this work, we propose a new Bayesian approach to interpret inconsistency estimators, which is based on the distribution of the physical inconsistency given the obtained value of the inconsistency estimator, i.e., $P({\rm{physical~ inconsistency}}|\rm{estimator})$, and which quantifies the statistical properties of the physical inconsistency. The key feature of this framework is to separate the concepts of physical inconsistency and data scatter; while the former manifests as the same parameter shift, the latter is random in each data realization. We will use the recently proposed index of inconsistency (IOI) (see Ref.\,\cite{WL2017a,WL2017b}) as a case study to illustrate the points made, but this Bayesian approach can be applied to inconsistency estimators in all generality. The new interpretation will also replace the original interpretation using Jeffrey's scales proposed in \cite{WL2017a}. To our knowledge, this is the first work in the literature that interprets the value of inconsistency measure based on the conditional probability distribution of physical inconsistency using Bayes' theorem. We will identify the underlying cause of the above mentioned ambiguity and show that the new Bayesian framework resolves such an ambiguity in a simple and natural way. We also apply the interpretation to current cosmological datasets and revisit some results. An easy-to-use public code is provided on the GitHub link given in the footnote\footnote{\href{https://github.com/WeikangLin/IOI.git}{https://github.com/WeikangLin/IOI.git}.\label{fn-IOI-git}}.

%%%% WL MODIFIED:
This paper is organized as follows. In Sec.\,\ref{sec:toy_realization_in_physical_inconsistency} we briefly discuss and use a toy model of two observations to introduce the \textit{level of physical inconsistency} $\lvb$. In Sec.\,\ref{sec:Bayesian_interpretation_IOI_in_scatters} we discuss the new Bayesian interpretation of IOI. Then in Sec.\,\ref{sec:application-Bayesian-interpretation} we apply IOI and the new Bayesian interpretation to examine the (in)consistencies between Planck versus the Cepheid-based local measurement, the Dark energy survey (DES) and the CMB temperature and polarization anisotropy from the Atacama Cosmology Telescope (ACT). The ambiguity in the quantification of inconsistency is discussed in Sec.\,\ref{sec:ambiguity}, where we show that it can be resolved using our Bayesian interpretation. Finally, we summarize and conclude in Sec.\,\ref{sec:IOI-conclusion-outlook}.

\section{Data scatters versus physical inconsistency}\label{sec:toy_realization_in_physical_inconsistency}
We first take a simple toy model of constraint comparison to introduce the definition of the level of physical inconsistency, $\lvb$. Imagine we are comparing the two constraints of a model with $N_{\rm{p}}$ parameters. Suppose that the model parameter vector $\bm{\lambda}$ is directly related to the observables $\bm{Q}$, (i.e., there is no transformation between $\bm{\lambda}$ and $\bm{Q}$) and that the two likelihoods are of Gaussian forms
\begin{equation}\label{eq:Q1-Q2-likelihood}
    \mathcal{L}_i(\bm{Q_i};\bm{\lambda})=\frac{1}{(2\pi)^{N_{\rm{p}}/2}\sqrt{|\bm{C_i}|}}\exp\left[ -\frac{1}{2}(\bm{Q_i}-\bm{\lambda})^T\bm{C_i}^{-1}(\bm{Q_i}-\bm{\lambda}) \right]\,.
\end{equation}

In this work, we focus on likelihoods that have a Gaussian functional form of the model parameters.\footnote{It should be pointed out that Gaussian likelihoods usually refer to those that have a Gaussian functional form of data.} We do so because (1) Gaussian cases are important cases in practice and as cosmological data get more and more powerful, likelihoods are more and more Gaussian. (2) Fundamental issues still need to be resolved in the quantification of inconsistency and Gaussian cases can be used as standard cases to illustrate the problem and to verify newly proposed inconsistency measures and interpretations.  As a result, special attention have been paid to Gaussian cases \cite{WL2017a,2018-Raveri-Hu,Handley-etal-2019-suspiciousness,2017-GB-Zhao-etal-BAO-surprise}.

Without loss of generality, let us assume that the true parameter vector is $\bm{\lambda}_{\rm{true}}=\bm{0}$ to simplify the expressions. We first assume that there is no physical inconsistency in the first experiment, i.e., the first data $\bm{Q_1}$ is a realization drawn from a likelihood with $\bm{\lambda}=\bm{\lambda}_{\rm{true}}$,
\begin{equation}\label{eq:Q1-true-likelihood}
    \mathcal{L}^{\rm{true}}_1(\bm{Q_1};\bm{\lambda_{\rm{true}}})=\frac{1}{(2\pi)^{N_{\rm{p}}/2}\sqrt{|\bm{C_1}|}}\exp\left[ -\frac{1}{2}\bm{Q_1}^T\bm{C_1}^{-1}\bm{Q_1} \right]\,.
\end{equation}
Let us assume the second observation is somehow biased, which manifests as a biased ``true parameter vector'' $\bm{\lambda_{\rm{biased}}}=\bm{\lambda_{\rm{true}}}+\bm{b}=\bm{b}$. Then the second data $\bm{Q_2}$ is a realization drawn from a likelihood with $\bm{\lambda}=\bm{b}$, that is,
\begin{equation}\label{eq:Q2-biased-likelihood}
    \mathcal{L}^{\rm{biased}}_2(\bm{Q_2};\bm{\lambda_{\rm{biased}}})=\frac{1}{(2\pi)^{N_{\rm{p}}/2}\sqrt{|\bm{C_2}|}}\exp\left[ -\frac{1}{2}(\bm{Q_2}-\bm{b})^T\bm{C_2}^{-1}(\bm{Q_2}-\bm{b}) \right]\,.
\end{equation}
After the two observations are made, we obtain two realizations, $\bm{Q_1}$ and $\bm{Q_2}$, drawn respectively from Eqs.\,\eqref{eq:Q1-true-likelihood} and \eqref{eq:Q2-biased-likelihood}. Then $\bm{Q_1}$ and $\bm{Q_2}$ are put in the two likelihoods in the form of Eq.\,\eqref{eq:Q1-Q2-likelihood} to perform a Bayesian analysis to constrain $\bm{\lambda}$.

We further assume the prior is weak, so that for this toy model the two posteriors equal the two likelihoods, i.e., $P_i(\bm{\lambda};\bm{Q_i})=\mathcal{L}_i(\bm{Q_i};\bm{\lambda})$. For our specific settings, the reported means of $\bm{\lambda}$ from the two experiments will be $\bm{Q_1}$ and $\bm{Q_2}$, respectively, and the reported covariance matrices are $\bm{C_1}$ and $\bm{C_2}$. It can be shown from Eqs.\,\eqref{eq:Q1-true-likelihood} and \eqref{eq:Q2-biased-likelihood} that the probability distribution of $\bm{\delta}\equiv\bm{Q_1}-\bm{Q_2}$ is,
\begin{equation}\label{eq:distribution_Q1-Q2}
    P(\bm{\delta};\bm{b})=\frac{1}{(2\pi)^{N_{\rm{p}}/2}\sqrt{|\bm{C_1}+\bm{C_2}|}}\exp\left[ -\frac{1}{2}(\bm{\delta}-\bm{b})^T(\bm{C_1}+\bm{C_2})^{-1}(\bm{\delta}-\bm{b}) \right]\,.
\end{equation} 
We define the level of physical inconsistency as
\begin{equation}\label{eq:physical-inconsistency-general}
    \lvb\equiv\sqrt{\bm{b}^{T}(\bm{C_1}+\bm{C_2})^{-1}\bm{b}}\,,
\end{equation}
which is also equal to $\sqrt{2\rm{IOI}}$ between the two constraints on $\bm{\lambda}$ in the presence of a physical inconsistency and in the absence of data scatter (i.e., when $\bm{Q_1}=\bm{0}$ and $\bm{Q_2}=\bm{b}$). The definition of such a level of physical inconsistency is motivated by the following points: (1) In the case when the bias is only in one parameter and there are no correlations between any parameters, Eq.\,\eqref{eq:physical-inconsistency-general} covers the usual one-parameter case; (2) there is always a continuous change to a general Gaussian case which allows it to become a one-parameter case, and it is Eq.\,\eqref{eq:physical-inconsistency-general} that can follow the continuous change and reduce to the one-parameter case.

Such a definition of the level of physical inconsistency $\lvb$ also includes the case where the inconsistency is caused by an incorrect underlying model. This is because an incorrect underlying model would manifest as different ``true'' parameter values for one type of observation than those for another type of observation. Therefore, any nonzero level of physical inconsistency $\lvb$ can be caused by biases in observations, an incorrect underlying model, or both. {
When $\lvb>1$, there is always one parameter (or a parameter combination) that suffers from some physical inconsistency with a normalized amplitude larger than unity. In such a case, conclusions based on the use of the two constraints become questionable. }

Now, data scatter mixes with the physical inconsistency and $\bm{\delta}$ will be different from $\bm{b}$. As a result, we cannot directly use
\begin{equation}\label{eq:IOI-with-delta}
    \sqrt{2\rm{IOI}}=\sqrt{\bm{\delta}^{T}(\bm{C_1}+\bm{C_2})^{-1}\bm{\delta}}\,,
\end{equation}
to quantify the physical inconsistency, because it will include the contribution from data scatter and the result is different from $\lvb$ in Eq.\,\eqref{eq:physical-inconsistency-general}. This does not necessarily mean that $\sqrt{2\rm{IOI}}$ will always be larger than $\lvb$, because it depends on the direction of data scatter. 

Now the question becomes, given an obtained value of $\sqrt{2\rm{IOI}}$, what statistical properties of $\lvb$ can we deduce? This will be discussed in the next section. 

Before we discuss our Bayesian interpretation, let's briefly recall the common practice that uses the probability to exceed (PTE) and the $n$-$\sigma$ notation of significance level. In the absence of any physical inconsistency, it has been shown that IOI, or some similar quantities, follows a $\chi$-square distribution ($f_{\chi^2}$) with a degree of freedom given by, e.g., the number of parameters considered \cite{2018-Raveri-Hu}. Then PTE - the probability of IOI being greater than the obtained value - is,
\begin{equation}
    \label{eq-convert-IOI-PTE}
    {\rm{PTE}}=\int_{2{\rm{IOI}}}^{\infty} f_{\chi^2}(z,N_{\rm{p}})dz=\gamma_{\rm{u}}(N/2,{\rm{IOI}})\,,
\end{equation}
where $\gamma_{\rm{u}}(s,x)\equiv\frac{1}{\Gamma(s)}\int_x^\infty t^{t-1}e^{-t}dt$ is the normalized upper incomplete gamma function. The common $n$-$\sigma$ notation of significance level is then associated with this PTE and is given by,
\begin{equation}\label{eq-significance-P-IOI}
    n\text{-}\sigma=\sqrt{2\gamma_{\rm{u}}^{\rm{inv}}(1/2,{\rm{PTE}})}\,,
\end{equation}
where $\gamma_{\rm{u}}^{\rm{inv}}$ is the inverse function of $\gamma_{\rm{u}}$.

\section{A Bayesian Interpretation of inconsistency measures: IOI case study}\label{sec:Bayesian_interpretation_IOI_in_scatters}
\subsection{The Bayesian framework}\label{sec:the-framework}
As mentioned above, the central question now is that given an obtained value of $\sqrt{2\rm{IOI}}$, what statistics of the level of physical inconsistency $\lvb$ can we deduce? The answer lies in obtaining the conditional probability distribution of the level of physical inconsistency $\lvb$, i.e., $P(\lvb|\sqrt{2\rm{IOI}})$. To do so, we use Bayes' theorem,
\begin{equation}\label{eq:P_b_IOI}
    P(\lvb|\sqrt{2\rm{IOI}})=\frac{P(\sqrt{2\rm{IOI}}|\lvb)\times P(\lvb)}{P(\sqrt{2\rm{IOI}})}\,,
\end{equation}
where $P(\lvb)$ is the prior distribution of $\lvb$ which we assume here to be flat for $\lvb\geq0$,\footnote{\label{fn-prior-beta}The adopted prior, $P(\lvb)$, may be improved by considering some comparison in the past between similar observations. Adopting other priors on $\lvb$ can be done by modifying the first of the four steps later discussed.} the likelihood $P(\sqrt{2\rm{IOI}}|\lvb)$ is the probability distribution of $\sqrt{2\rm{IOI}}$ given $\lvb$, the denominator $P(\sqrt{2\rm{IOI}})$ only serves as a normalization factor, and  $P(\lvb|\sqrt{2\rm{IOI}})$ is the posterior.

\begin{figure}[h]
    \centering
    \includegraphics[width=0.75\textwidth]{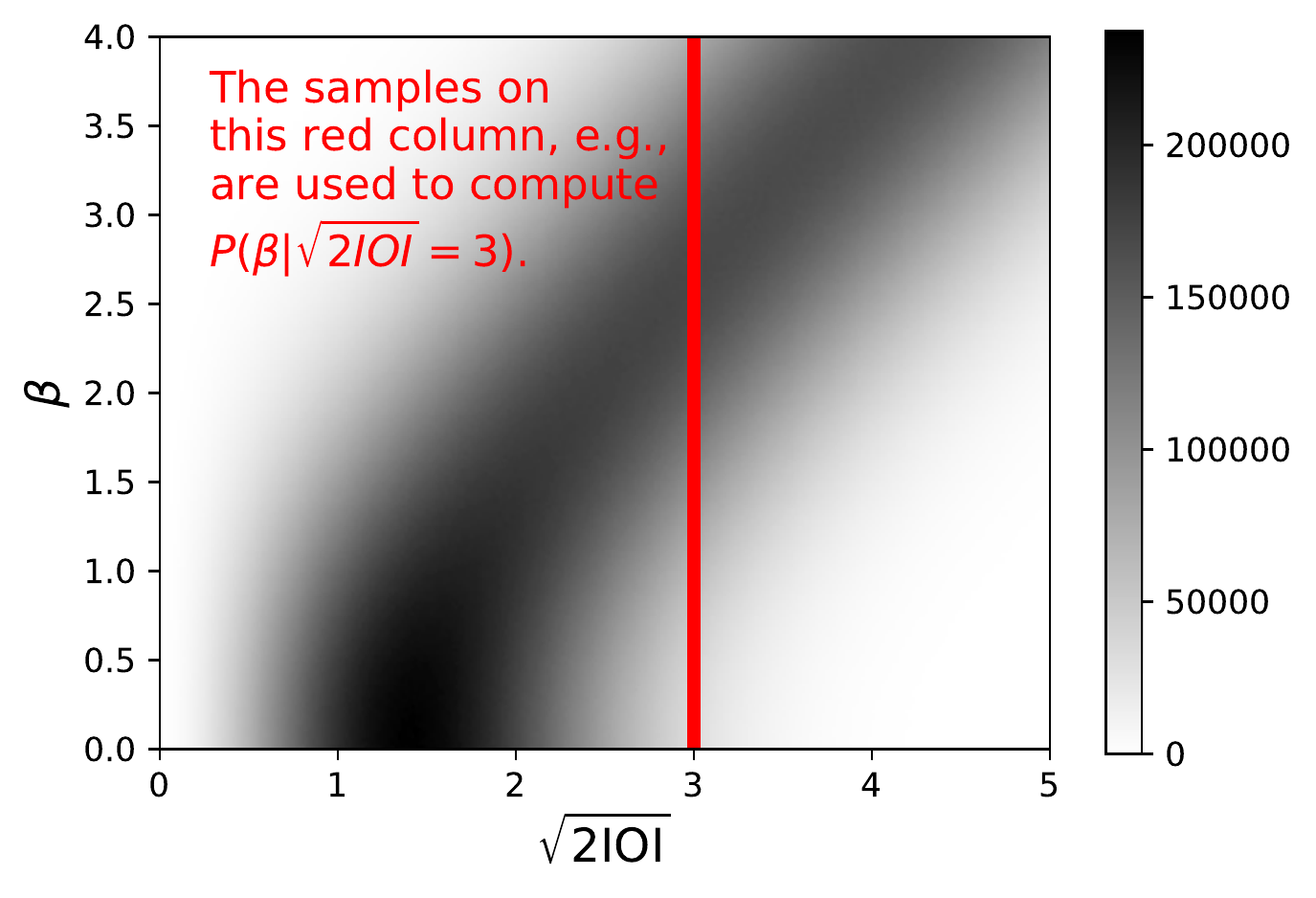}
    \caption{An illustrative example of the histogram obtained by the simulations described in section \ref{sec:the-framework} with $N_{\rm{p}}=3$. The darkness represents the number of samples on the grid points. The red column shows a column of samples that is used to compute $P(\lvb|\sqrt{2\rm{IOI}}=3)$.}
    \label{fig:simulation_example}
\end{figure}

{ First, we verify with simulations that for the same value of $\lvb$, the likelihood $P(\sqrt{2\rm{IOI}}|\lvb)$ is the same for a general situation where $\bm{b}$, $\bm{C_1}$ and $\bm{C_2}$ are arbitrary. So, for the same prior $P(\lvb)$, the posterior $P(\lvb|\sqrt{2\rm{IOI}})$ is the same according to Eq.\,\eqref{eq:P_b_IOI} regardless of the forms of $\bm{b}$, $\bm{C_1}$ and $\bm{C_2}$. Therefore, we simplify the calculation by taking $\bm{b}$ to be along the direction of one parameter and that $\bm{C_1}+\bm{C_2}={\rm{diag}}(1,1,1,1,\cdots)$, so that $\bm{b}=(\lvb,0,0,0,\cdots)$ with no loss of generality. 

Then, to evaluate $P(\lvb|\sqrt{2\rm{IOI}})$, we perform a likelihood-free inference and perform simulations to obtain two-dimensional histograms in the $\sqrt{2\rm{IOI}}$-$\lvb$ space that allow one to infer both $P(\sqrt{2\rm{IOI}}|\lvb)$ and $P(\lvb|\sqrt{2\rm{IOI}})$ as follows. } We generate a two-dimensional grid in the $\lvb$-$\sqrt{2\rm{IOI}}$ space, spanning from $0\leq \lvb\leq 10$ and $0\leq \sqrt{2\rm{IOI}}\leq 13$. This range of $\lvb$ is enough for most practical cases, but the framework can be extended to include even larger values of $\lvb$. The range of $\sqrt{2\rm{IOI}}$ needs to be bigger than that of $\lvb$ for a good sampling of the likelihood $P(\sqrt{2\rm{IOI}}|\lvb)$.  Next, we choose a parameter number ($N_{\rm{p}}$) and repeat the following four steps $10^{10}$ times with each chosen $N_{\rm{p}}$:
\begin{enumerate}
    \item Randomly but uniformly draw one value of $\lvb_{\rm{grid}}$ on the grids\footnoteref{fn-prior-beta};
    \item Draw one realization of the two observations according to likelihoods \eqref{eq:Q1-true-likelihood} and \eqref{eq:Q2-biased-likelihood};
    \item Calculate the corresponding $\sqrt{2\rm{IOI}}$ according to Eq.\,\eqref{eq:IOI-with-delta} with the drawn data; we denote it as $\sqrt{2\rm{IOI}_{\rm{real.}}}$;
    \item Add a count to the grid point $(\lvb_{\rm{grid}}, \sqrt{2\rm{IOI}_{\rm{grid}}})$ if the value of $\sqrt{2\rm{IOI}_{\rm{real.}}}$ falls into the interval $\big(\sqrt{2\rm{IOI}_{\rm{grid}}}-h,~\sqrt{2\rm{IOI}_{\rm{grid}}}~\big]$, i.e.,  if $\sqrt{2\rm{IOI}_{\rm{grid}}}-h< \sqrt{2\rm{IOI}_{\rm{real.}}} \leq \sqrt{2\rm{IOI}_{\rm{grid}}}$.
\end{enumerate}
After the above steps, we obtain a two-dimensional histogram in the $\lvb$-$\sqrt{2\rm{IOI}}$ space for each $N_{\rm{p}}$. Figure \ref{fig:simulation_example} shows an example of such histograms with $N_{\rm{p}}=3$. 

{For a column with the same value of $\sqrt{2\rm{IOI_{grid}}}$ in a histogram, such as the one  in red in Figure \ref{fig:simulation_example}, the value of $P(\lvb|\sqrt{2\rm{IOI_{grid}}})$ is proportional to the number of the samples on the grid, $n(\lvb_{\rm{grid}},\,\sqrt{2\rm{IOI_{grid}}})$. Then, $P(\lvb|\sqrt{2\rm{IOI}})$ can be obtained by normalizing the number distribution for each constant-$\sqrt{2\rm{IOI}}$ column, that is,
\begin{equation}\label{eq:obtain-P-by-normalization}
    P(\lvb_{\rm{grid}}|\sqrt{2\rm{IOI_{grid}}})=\frac{n(\lvb_{\rm{grid}},\,\sqrt{2\rm{IOI_{grid}}})}{\sum\limits_{\lvb_{\rm{grid}}}n(\lvb_{\rm{grid}},\,\sqrt{2\rm{IOI_{grid}}})}\,.
\end{equation}
Note that now we do the above only for $0\leq \sqrt{2\rm{IOI}}\leq7$. This is because for the posterior $P(\lvb|\sqrt{2\rm{IOI}})$ the range of $\lvb$ now needs to be larger than that of $\sqrt{2\rm{IOI}}$ to capture the tail of the distribution at large $\lvb$. The value of $P(\lvb|\sqrt{2\rm{IOI}})$ for general $\lvb$ and $\sqrt{2\rm{IOI}}$ values can then be obtained by interpolation.}

\subsection{Two summary statistics for the level of physical inconsistency}
Now that $P(\lvb|\sqrt{2\rm{IOI}})$ \footnote{A bonus point: by using the conditional probability of $\lvb$, it does not matter whether we use IOI or $\sqrt{2\rm{IOI}}$. This is because now IOI and $\sqrt{2\rm{IOI}}$ become parameters and $P(\lvb|\rm{IOI})=P(\lvb|\sqrt{2\rm{IOI}})$. So, we use $P(\lvb|\rm{IOI})$ and $P(\lvb|\sqrt{2\rm{IOI}})$ interchangeably.} is evaluated, we can obtain some statistical properties regarding the level of physical inconsistency. There are two important summary statistics to calculate, which correspond to two different aspects in the quantification of inconsistency: 
\begin{description}%[align=left,leftmargin=5pt]
\item[Question 1:] What is the probability for the level of physical inconsistency to be greater than unity, i.e., $P(\lvb>1|\sqrt{2\rm{IOI}})$?
\item[Question 2:] What are the $68\%$- and $95\%$-percentile ranges of the level of physical inconsistency $\lvb$?
\end{description}

The first question is about the level of confidence in the presence of a non-negligible physical inconsistency. The commonly adopted $n$-$\sigma$ notation is actually also a quantity trying to address this question, but in an indirect way; see the next subsection for discussion. But this first question does not represent the extent to which, e.g., an observation suffers from a bias. Such an extent is quantified by $\lvb$, which leads to our second question. 

The second question is about the (normalized) magnitude of the physical inconsistency $\lvb$. It is the value of $\lvb$ that represents the extent to which a bias or an invalid model affects one or both constraints. The simplest example is when only one parameter in one constraint is biased. Therefore, this second question is more important, and if one would like to rank the intensity of an inconsistency it is the value of $\lvb$ that one should refer to.

The above two questions represent two different aspects in the quantification of inconsistency. As we will discuss in detail in Sec.\,\ref{sec:ambiguity} and demonstrate with real examples in Sec.\,\ref{sec:application-Bayesian-interpretation}, these two aspects behave differently as the number of model parameters increases. In the literature, other inconsistency measures dealing with data scatter are not Bayesian approaches. Fundamentally, other measures are not treating physical inconsistency and data scatter separately because they are based on the probability of a measure in the absence of any physical inconsistency. As a consequence, other measures are not able to distinguish the two aspects mentioned above, which leads to a long-standing ambiguity in the quantification of inconsistency in multiple-parameter cases; see Sec.\,\ref{sec:application-Bayesian-interpretation} and Sec.\,\ref{sec:ambiguity} for detailed discussion. 

In the following subsections, we will discuss these two summary statistics, respectively.

\subsection{The threshold criterion for the presence of some physical inconsistency}\label{sec:probabilityy-of-presence}
\begin{figure}[tbp]
    \centering
    \includegraphics[width=0.65\textwidth]{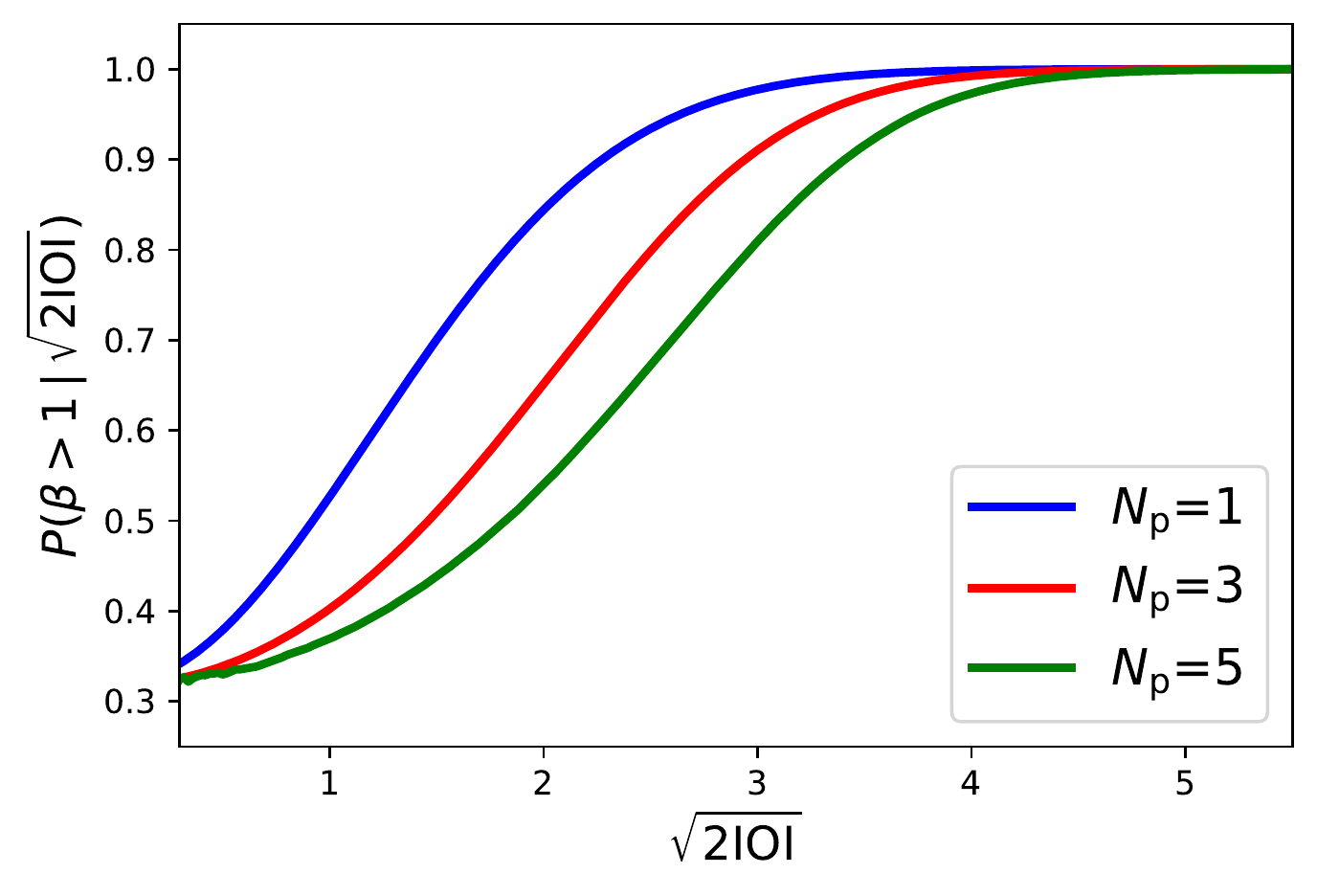}
    \caption{The probability for the physical inconsistency to reach the threshold ($\lvb>1$) given an obtained value of $\sqrt{2\rm{IOI}}$, $P(\lvb>1|\sqrt{2\rm{IOI}})$, for different parameter dimensions. For a given value of $\sqrt{2\rm{IOI}}$, such a probability is larger with a larger number of parameters.}
    \label{fig:P_of_b1}
\end{figure}

From the first question above, we can see whether there is some physical inconsistency that reaches the threshold over which the inconsistency needs attention. As mentioned, the commonly adopted $n$-$\sigma$ notation actually tries to address this question indirectly. The usual $n$-$\sigma$ value is associated with a PTE that quantifies the probability of that estimator being equal to or larger than the obtained value when there is no physical inconsistency, i.e., $P(\rm{estimator}>\rm{estimator~obtained}|\lvb=0)$. Our first question, on the other hand, is based on the conditional probability distribution of the level of physical inconsistency given an obtained value of inconsistency estimator (in this case IOI), i.e., $P(\lvb|\sqrt{2\rm{IOI}})$, and directly quantifies the probability of physical inconsistency large enough to be further considered, i.e., $P(\lvb>1|\sqrt{2\rm{IOI}})$. Therefore, compared to the usual $n$-$\sigma$ value, our first question address a more direct and desirable statistical property of the physical inconsistency. The probability $P(\lvb>1|\sqrt{2\rm{IOI}})$ is calculated as,
\begin{equation}\label{eq:P_b1_givenIOI}
    P(\lvb>1|\sqrt{2\rm{IOI}})=\int_1^\infty d\lvb P(\lvb|\sqrt{2\rm{IOI}})\,.
\end{equation}
We show some results of $P(\lvb>1|\sqrt{2\rm{IOI}})$ in Figure \ref{fig:P_of_b1}. 

{The reason for the choice of $\lvb>1$ is as follows. When $\lvb\ll1$, any bias or inconsistency due to an invalid model will be  negligible because its magnitude is much less than $\sqrt{\sigma_{1,{\rm{a}}}^2+\sigma_{2,{\rm{a}}}^2}$, where $\sigma_{1,{\rm{a}}}$ and $\sigma_{2,{\rm{a}}}$ (``a'' for ``any'') are the uncertainties of any model parameter or any linear combination of model parameters obtained from the two posterior distributions of concern. In other words, we need at least $\lvb\sim1$ so that the physical inconsistency can be resolved with the current level of uncertainty. Therefore, it is well justified to calculate $P(\lvb>1|\sqrt{2\rm{IOI}})$. Note that it is incorrect both mathematically and physically to ask for something like the probability ratio of $\lvb=0$ versus $\lvb\neq0$. Mathematically, since $P(\lvb|\sqrt{2\rm{IOI}})$ is a distribution of a continuous variable $\lvb$ the probability of $\lvb=0$ is zero. Physically, the magnitude of physical inconsistency cannot be completely zero, because the real situation is always not ideal. Even if there is no beyond-the-standard physics, there is always some unaccounted-for systematic error to some extent. So, the good case is when $\lvb\ll1$ rather than $\lvb=0$.}

Now, with $P(\lvb>1|\sqrt{2\rm{IOI}})$, we can use the usual $\alpha$ level as a criterion to judge whether there is a physical inconsistency reaching the threshold. For example, we can set $\alpha=0.15$, and require $P(\lvb>1|\sqrt{2\rm{IOI}})>(1-0.15)=85\%$ for such a threshold. For a given $\alpha$ level, there is a threshold value, $\rm{IOI}_{\rm{th}}$, that when $\rm{IOI}>\rm{IOI}_{\rm{th}}$ there is at least a probability of $(1-\alpha)$ for the level of physical inconsistency to be $\lvb>1$. We show in Table \ref{tab:IOI_alpha_values} the threshold $\rm{IOI}_{\rm{th}}$ and $\sqrt{2\rm{IOI}_{\rm{th}}}$ corresponding to some commonly chosen values of $\alpha$ with different $N_{\rm{p}}$'s. We suggest to set $\alpha=0.15$, because $\sqrt{2\rm{IOI}_{\rm{th}}}$ in the one-parameter case corresponds to the usual $2$-$\sigma$ confidence level (see Table \ref{tab:IOI_alpha_values}) around which the inconsistency begins to cause attention. This should not be taken as a circular definition, but as a purpose of calibration. After all, in a one-parameter case the commonly adopted $n$-$\sigma$ notation works well and it is only necessary to use our Bayesian interpretation in higher dimensional cases; see Sec.\,\ref{sec:ambiguity} for a discussion. 

\begin{table}[tbp]
\renewcommand{\arraystretch}{1.3}
    \centering
    \begin{tabular}{l|c|cccccc}
    \hline\hline
        & $N_{\rm{p}}$ & 1 & 2 & 3 & 4 & 5 & 6  \\
    \hline
    \multirow{3}{*}{$\alpha=0.15$}    & $\rm{IOI}_{\rm{th}}$ & 2.0 & 2.9 & 3.6 & 4.3 & 5.0 & 5.6 \\
    \cline{2-8}
    & $\sqrt{2\rm{IOI}_{\rm{th}}}$ & 2.0 & 2.4 & 2.7 & 2.9 & 3.2 & 3.4\\
    \cline{2-8}
    & Range of $\lvb$ at IOI$_{\rm{th}}$ & $2.0^{+1.0}_{-1.0}$ & $2.1^{+1.1}_{-1.1}$ & $2.2^{+1.1}_{-1.2}$ & $2.2^{+1.1}_{-1.2}$ & $2.3^{+1.2}_{-1.3}$ & $2.3^{+1.2}_{-1.3}$ \\
    \hline
    \multirow{3}{*}{$\alpha=0.1$}    & $\rm{IOI}_{\rm{th}}$ & 2.6 & 3.5 & 4.3 & 5.1 & 5.8 & 6.5 \\
    \cline{2-8}
        & $\sqrt{2\rm{IOI}_{\rm{th}}}$ & 2.3 & 2.7 & 2.9 & 3.2 & 3.4 & 3.6\\
    \cline{2-8}
    & Range of $\lvb$ at IOI$_{\rm{th}}$ & $2.3^{+1.0}_{-1.0}$ & $2.4^{+1.0}_{-1.1}$ & $2.5^{+1.1}_{-1.2}$ & $2.6^{+1.1}_{-1.2}$ & $2.6^{+1.1}_{-1.3}$ & $2.7^{+1.2}_{-1.3}$ \\
     \hline\hline
    \end{tabular}
    \caption{Values of $\rm{IOI}_{\rm{th}}$ and $\sqrt{2\rm{IOI}_{\rm{th}}}$ as well as the ranges of $\lvb$ at IOI$_{\rm{th}}$ are also listed corresponding to $P(\lvb>1|\sqrt{2\rm{IOI_{\rm{th}}}})=(1-\alpha$) in different parameter dimensions ($N_{\rm{p}}$) for a commonly chosen $\alpha=0.1$. The value of IOI needs to be larger than the $\rm{IOI}_{\rm{th}}$ corresponding to the chosen $\alpha$ level and the number of (common) model parameters so that we can conclude that there is likely some physical inconsistency that reaches the threshold (i.e., $\lvb>1$). This table is provided for the convenience of IOI users, but $P(\lvb>1|\sqrt{2\rm{IOI_{\rm{th}}}})$ can be obtained for an arbitrary value of IOI with our code provided (see footnote \ref{fn-IOI-git}). When IOI is greater than $\rm{IOI}_{\rm{th}}$, we then need to find the most probable ranges of the level of physical inconsistency. We suggest to use $\alpha=0.15$; see the text for discussion.}
    \label{tab:IOI_alpha_values}
    \renewcommand{\arraystretch}{1}
\end{table}

\subsection{Inference of the range of the actual level of physical inconsistency}\label{sec:inference-actual-level}
\begin{figure}[tbp]
    \centering
    \includegraphics[width=\textwidth]{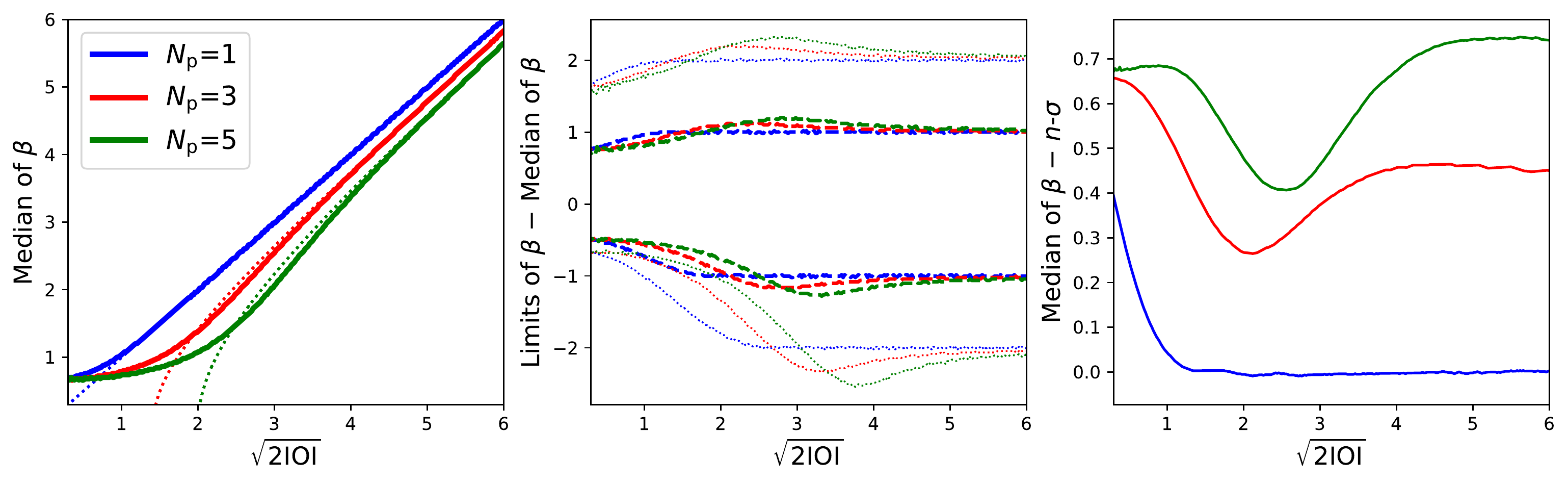}
    \caption{Median statistics for the level of physical inconsistency $\lvb$ obtained from $P(\lvb|\sqrt{2\rm{IOI}})$. In all panels, different colors represent $N_{\rm{p}}=1,\,3,\,5$, as in the left panel. Left: the median of $\lvb$ as a function of $\sqrt{2\rm{IOI}}$ shown in solid lines for $N_{\rm{p}}=1,2,3$. For comparison, we also show in dotted lines the asymptotic value of $\lvb$, $\sqrt{2\rm{IOI}-(N_{\rm{p}}-1)}$, at large values of IOI, i.e., when $\sqrt{2\rm{IOI}}\gtrsim\sqrt{N_{\rm{p}}}$. Middle: the ranges of $\lvb$ subtracted from its median. The thick, dashed curves correspond to the $68\%$ upper (shown above $0$) and lower (shown below $0$) limits as a function of $\sqrt{2\rm{IOI}}$. The $68\%$ range of $\lvb$ is about $\pm1$ off its median. The thin, dotted curves correspond to the $95\%$ limits. Right: the difference between the median of $\lvb$ and the usual $n$-$\sigma$ value as a function of $\sqrt{2\rm{IOI}}$. The $n$-$\sigma$ value is smaller than the median of $\lvb$ when there is more than one parameter. }
    \label{fig:IOI_to_b}
\end{figure}

To answer the second question, we use the standard median statistics. Then, the median of $\lvb$ is defined as the root of the following equation,
\begin{equation}\label{eq:def-median-beta}
    \int_0^{\lvb_{\rm{med}}} d\lvb P(\lvb|\sqrt{2\rm{IOI}})=50\%\,.
\end{equation}
The lower and upper limits of the $68\%$-percentile range of $\lvb$ are defined as 
\begin{equation}\label{eq:lower-upper-68-beta}
    \int_0^{\lvb_{\rm{low}}^{68\%}\,(\rm{or}~\lvb_{\rm{up}}^{68\%})} d\lvb P(\lvb|\sqrt{2\rm{IOI}})=16\%~({\rm{or}}~84\%)\,.
\end{equation}
The lower and upper limits of the $95\%$-percentile range of $\lvb$ are defined in a similar way. In the left panel of Figure \ref{fig:IOI_to_b}, we show the median of the level of physical inconsistency, $\lvb_{\rm{med}}$, as a function of the obtained value of $\sqrt{2\rm{IOI}}$ with three different parameter numbers, $N_{\rm{p}}=1,3,5$; see the solid curves. We also show in dashed lines the asymptotic value of $\lvb_{\rm{med}}$,
\begin{equation}\label{eq:asymptotic-value-beta-median}
    \lim_{{\rm{IOI}}\rightarrow\infty}\lvb_{\rm{med}}=\sqrt{2\rm{IOI}-(N_{\rm{p}}-1)}\,,
\end{equation}
which matches the $\lvb_{\rm{med}}$ very well at high enough values of $\sqrt{2\rm{IOI}}$, i.e., $\sqrt{2\rm{IOI}}\gtrsim\sqrt{N_{\rm{p}}}$. In the middle panel, we show the $68\%$- and $95\%$-percentile ranges of $\lvb$ as a function of $\sqrt{2\rm{IOI}}$. As we can see, the $68\%$-percentile range of $\lvb$ is roughly $\pm1$ away from the median of $\lvb$ for a large enough $\sqrt{2\rm{IOI}}$. With these above, a quick estimate of the $68\%$-percentile range of $\lvb$ is $\sqrt{2\rm{IOI}-(N_{\rm{p}}-1)}\pm1$, which works for $\sqrt{2\rm{IOI}}\gtrsim\sqrt{N_{\rm{p}}}$ and is better for a larger $\sqrt{2\rm{IOI}}$. The $95\%$-percentile range of $\lvb$ is also quite regular and is roughly $\pm2$ away from the median of $\lvb$, but requires a larger value of $\sqrt{2\rm{IOI}}$ for the quick estimate to match the actual $95\%$-percentile range of $\lvb$. Note that in the one-parameter case, the median of $\lvb$ equals the usual significance level if $\lvb\gtrsim1$.

Thus, our proposed Bayesian interpretation of IOI consists of calculating the two summary statistics of the level of physical inconsistency dependent on the obtained value of IOI, and as we described above, the process is twofold. These two statistics work in tandem. The first allows one to estimate whether the physical inconsistency reaches the threshold to be further considered. Next, when $P(\lvb>1|\sqrt{2\rm{IOI}})$ is high enough, one needs to use the second statistic to determine the actual ranges of the level of that physical inconsistency. We provide the code to calculate the relevant summary statistics of the level of physical inconsistency based on our new Bayesian interpretation of IOI at the link indicated in the footnote \ref{fn-IOI-git}.

It is important to note that in multiple-parameter cases one should not take the non-Bayesian $n$-$\sigma$ value as the level of physical inconsistency as it would tend to underestimate the value of $\lvb$. This can be seen in the right panel of Figure \ref{fig:IOI_to_b} where we plot the difference between the median of $\lvb$ and the $n$-$\sigma$ value as a function of $\sqrt{2\rm{IOI}}$. In one parameter cases, the difference is essentially zero when there is a noticeable physical inconsistency. However, when $N_{\rm{p}}>1$ the $n$-$\sigma$ value is generally smaller than the median of $\lvb$. The difference is more pronounced with more parameters. We will demonstrate with real data in Sec.\,\ref{sec:application-Bayesian-interpretation} and will discuss the underlying reason behind this issue in Sec.\,\ref{sec:ambiguity}.

\subsection{A ranking scheme of the level of physical inconsistency}\label{sec:guiding-ranking-scheme}
{The level of physical inconsistency $\lvb$ is already presented as a normalized magnitude. Therefore, one can take its face value as an interpretation of the strength of inconsistency due to a bias or an invalid underlying model. In practice however, one often, if not always, imposes some empirical interpretation on the values of inconsistency measures whether or not such values are obtained in a statistical manner. Indeed, even in one-parameter cases, where inconsistency can be presented with the non-Bayesian $n$-$\sigma$ value, some empirical interpretation is often adopted. One may argue that the $n$-$\sigma$ value is associated with the PTE of some inconsistency measure which has some definite statistical meaning. But it is empirical, e.g., to only consider an inconsistency larger than $2$-$\sigma$ or $3$-$\sigma$ as non-negligible. A recent example of empirical interpretation (for one-parameter cases) can be found in Ref.\,\cite{2019-Verde-Treu-Riess}, i.e., $2$-$\sigma$=curiosity; $3$-$\sigma$=tension; $4$-$\sigma$=discrepancy; $5$-$\sigma$=crisis.

\begin{table}[tbp]
\renewcommand{\arraystretch}{1}
\end{table}
\begin{table}[t]
\begin{threeparttable}
    \centering
    \renewcommand{\arraystretch}{1.2}
    \begin{tabular}{p{0.2\textwidth}|p{0.28\textwidth}|p{0.425\textwidth}}
    \hline\hline
        \multicolumn{2}{c|}{Cases/situations} & Ranking of inconsistency \\
        \hline  
        \hline
        \multicolumn{2}{p{.5\textwidth}|}{IOI$<$IOI$_{\rm{th}}$: value of IOI is below the threshold listed in Table \ref{tab:IOI_alpha_values}. We suggest to use $\alpha=0.15$.} & No strong indication for the presence of a physical inconsistency that exceeds the threshold. \\
        \hline
        \multirow{5}{0.2\textwidth}{IOI$>$IOI$_{\rm{th}}$  and when the $68\%$ range of $\lvb$ falls into the cases on the right} &is all in zone 1. & Moderate inconsistency \\
        \cline{2-3}
         & spans both zone 1 and zone 2. & Moderate-to-strong inconsistency\\
        \cline{2-3}
         & spans zones 1, 2 and 3. & Strong inconsistency \\
        \cline{2-3}
         & spans both zone 2 and zone3. & Strong-to-very strong inconsistency \\
        \cline{2-3}
         & is all in zone 3. & Very strong inconsistency\\
    \hline\hline
    \end{tabular}
    \caption{A summary of the ranking scheme of the physical inconsistency in our Bayesian framework. This process is twofold: Step-1: IOI needs to be greater than the threshold IOI$_{\rm{th}}$ listed in Table \ref{tab:IOI_alpha_values}, indicating that there likely exists some physical inconsistency. Step-2: Then the $68\%$-percentile range of $\lvb$ is compared to three zones: zone 1 = (0,\,4), zone 2 = [4,\,5] and zone 3 = (5,\,$+\infty$) and is ranked accordingly. Examples of physical inconsistency in different ranks are shown in Figure \ref{fig:examples_of_beta_ranges}.}
    \label{table:new-scale2}
\end{threeparttable}
\end{table}

\begin{figure}[t!]
    \centering
    \includegraphics[width=\textwidth]{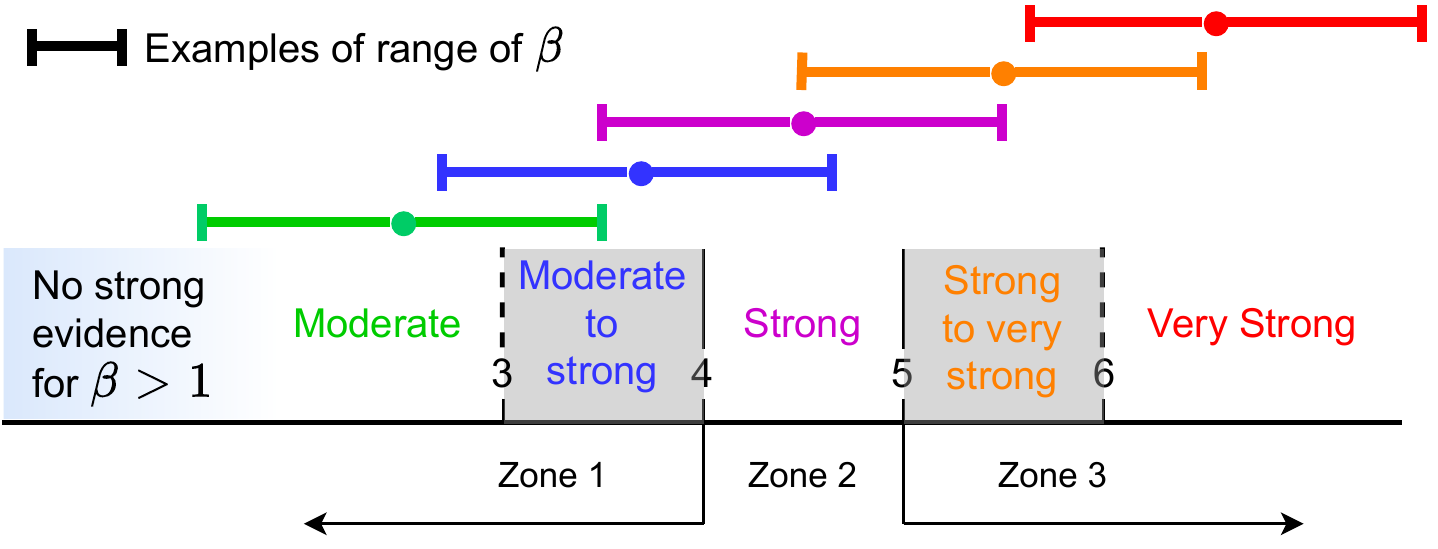}
    \caption[scales for IOI]{\label{tab:new-scale}
    Examples of the $68\%$-percentile ranges of $\lvb$ in different ranks. }
    \label{fig:examples_of_beta_ranges}
\end{figure}

As discussed in the previous sections, in multiple-parameter cases, it is the level of the physical inconsistency $\lvb$ instead of the commonly adopted $n$-$\sigma$ value that should be interpreted as the strength of inconsistency due to a bias or an invalid underlying model. In other words, if some empirical interpretation is imposed, it is $\lvb$ instead of the commonly adopted $n$-$\sigma$ value that such an empirical interpretation should be associated with. }

Therefore, in this subsection we suggest a guiding scale to interpret the level of physical inconsistency. We adopt a similar scale suggested in Ref.\,\cite{2019-Verde-Treu-Riess} about the level of tension with some modifications to the terminology and the ranking scheme. We note that, instead of a single value, $\lvb$ is now reported as the $68\%$-percentile range (as well as the $95\%$-percentile range). Therefore, our ranking scheme is made with respect to the range of $\lvb$ instead of a single value of it. Our ranking scheme is described as follows. First, the probability for $\lvb>1$ is compared to ($1-\alpha$), where we suggest to use $\alpha=0.15$. If $P(\lvb>1|\sqrt{2\rm{IOI}})<(1-\alpha)$, we conclude that there is no strong evidence for the presence of a physical inconsistency that exceeds the threshold. Otherwise the level of physical inconsistency needs further attention and the $68\%$-percentile range of $\lvb$ is compared to three zones: zone 1 = (0,\,4), zone 2 = [4,\,5] and zone 3 = (5,\,$+\infty$). When the range of $\lvb$ falls entirely in zone 1, we rank the physical inconsistency as a moderate inconsistency. When it overlaps with zone 1 and zone 2, it is ranked as a moderate-to-strong inconsistency. When it overlaps all three zones, it is ranked as a strong inconsistency. When it overlaps with zone 2 and zone 3, it is ranked as a strong-to-very-strong inconsistency. And finally, when it entirely falls in zone 3, it is ranked as a very strong inconsistency. We summarize our ranking scheme in Table \ref{table:new-scale2}.

Roughly speaking, according to the above ranking scheme, the median of $\lvb$ is $\lesssim3$ for a moderate inconsistency, between $\sim3$ and $\sim4$ for a moderate-to-strong inconsistency, between $\sim4$ and $\sim5$ for a strong inconsistency, between $\sim5$ and $\sim6$ for a strong-to-very-strong inconsistency, and $\gtrsim6$ for a very strong inconsistency. Examples of ranges of $\lvb$ in different ranks are shown in Figure \ref{fig:examples_of_beta_ranges}. Thus, in terms of $\lvb_{\rm{med}}$, this ranking scheme has a similar gradient as that suggested in Ref.\,\cite{2019-Verde-Treu-Riess} for the $n$-$\sigma$ notation in one-parameter cases, but in multiple-parameter cases it is our ranking scheme that should be adopted. Also note that we did not rank the inconsistency directly using the median of $\lvb$ because it is the range of $\lvb$ that preserves the same percentage in cases with different parameter numbers. But for the same range of $\lvb$, the value of $\lvb_{\rm{med}}$ is somewhat different for different parameter numbers.

\section{Applying the new Bayesian interpretation to revisit the (in)consistency between Planck versus the local measurement, DES, ACT and WMAP}\label{sec:application-Bayesian-interpretation}
In this section, we apply our Bayesian interpretation to compare some real data. One important point that we will show is that this Bayesian interpretation of IOI is insensitive to degrees of freedom irrelevant to the inconsistency. In this work we set $\alpha=0.15$ as suggested in Sec.\,\ref{sec:guiding-ranking-scheme}. We would also like to remind the reader that, when using IOI, it is only the common parameters of the two constraints that need to be considered \cite{WL2017a}.

\subsection{Case I: Planck versus the local measurement}\label{sec:IOI-Bayesian-H0}
In the first example, we apply IOI and our new Bayesian interpretation to revisit the (in)consistency between the Planck 2018 baseline temperature and polarization constraint (Planck) \cite[chains obtained from \href{https://pla.esac.esa.int}{https://pla.esac.esa.int}]{2018-Planck-cosmo-params} and the Cepheid-based local measurement \cite{2021-Riess-etal-H0}. The results are shown in Table \ref{table:planck-H0}. For the case of Planck 18 vs. the Cepheid-based local measurement ($H_0$ only), we have $\rm{IOI}=8.5$ with $N_{\rm{p}}=1$. This is a one-parameter case. In principle, we can use the usual significance level (which also equals $\sqrt{2\rm{IOI}}=4.1$) to interpret such an IOI value. But to be consistent with multiple-parameter cases, we will now use our newly provided Bayesian interpretation.  This high IOI value exceeds the thresholds listed in Table \ref{tab:IOI_alpha_values}, signaling the presence of a physical inconsistency that is caused by some unaccounted-for systematic effects or an invalid underlying model. In fact, the probability for $\lvb>1$ is $99.9\%$. The $68\%$-percentile range of the actual level of physical inconsistency is $\lvb=4.1^{+1.0}_{-1.0}$. This is a strong inconsistency according to the ranking scheme discussed in Sec.\,\ref{sec:guiding-ranking-scheme}.  

It is illustrative to extend this comparison between Planck and the Cepheid-based local measurement to a multiple-parameter case to see that our Bayesian interpretation is insensitive to the number of parameters in the underlying model which are irrelevant to the tension. We first extend this case to compare Planck 18 vs. local $H_0$+Pantheon in the $H_0$-$\Omega_{\rm{m}}$ space where we have $\rm{IOI}=8.6$. This is now a two-parameter case. The constraint on $\Omega_{\rm{m}}$ between Planck and Pantheon is consistent with each other, so expanding the comparison into Planck 18 vs. local $H_0$+Pantheon in the $H_0$-$\Omega_{\rm{m}}$ space should give us a similar result as the comparison of Planck 18 vs. local $H_0$ in the $H_0$-only space. Indeed, the IOI also exceeds the threshold values [see the $N_{\rm{p}}=2$ column in Table \ref{tab:IOI_alpha_values}], again convincingly signaling the presence of some physical inconsistency that needs attention. As expected, the $68\%$-percentile range of the level of physical inconsistency is $\lvb=4.0^{+1.0}_{-1.0}$, which is a similar range as that in the one-parameter case. 

\begin{table}[tbp]
\centering
\renewcommand{\arraystretch}{1.2}
\begin{tabular}{r|ccc}
\hline\hline
 & $H_0$, $N_{\rm{p}}=1$ & $H_0$-$\Omega_{\rm{m}}$, $N_{\rm{p}}=2$ & $H_0$-$\Omega_{\rm{m}}$-$\Omega_{\rm{b}}h^2$, $N_{\rm{p}}=3$  \\
 \hline
 $\sqrt{2\rm{IOI}}$ & $4.1>\sqrt{2\rm{IOI}_{\rm{th}}}$ & $4.1>\sqrt{2\rm{IOI}_{\rm{th}}}$ & $4.2>\sqrt{2\rm{IOI}_{\rm{th}}}$ \\
 Range of $\lvb$ & $4.1^{+1.0}_{-1.0}$ & $4.0^{+1.0}_{-1.0}$ & $3.9^{+1.0}_{-1.0}$ \\
 The commonly adopted $n$-$\sigma$ & $4.1$ & $3.7$ & $3.4$ \\
\hline\hline
\end{tabular}
\caption[Planck vs Local measurement table]{\label{table:planck-H0}Applying IOI and our new Bayesian interpretation to investigate the (in)consistency between Planck versus Cepheid-based local measurement (+Pantheon +BBN). The number of parameters increases from $N_{\rm{p}}=1$ ($H_0$) to $N_{\rm{p}}=3$ ($H_0$, $\Omega_{\rm{m}}$, $\Omega_{\rm{b}}h^2$) as we successively include more parameters in the comparison. The tension is mainly in the $H_0$ direction, so when we include in the comparison the Pantheon constraint on $\Omega_{\rm{m}}$ and the BBN constraint on $\Omega_{\rm{b}}h^2$, the conclusion about inconsistency should not change. Indeed, we can see that $\lvb$ remains in about the same range and indicates a strong inconsistency when $N_{\rm{p}}$ increases. On the contrary, the commonly adopted $n$-$\sigma$ notation of significance level drops as number of parameters increases. }
\end{table}

Next we extend the situation to compare Planck vs. local $H_0$+Pantheon+BBN in the $H_0$, $\Omega_{\rm{m}}$ and $\Omega_{\rm{b}}h^2$ space, where we have $\rm{IOI}=8.8$. Again, Planck and BBN are consistent in the $\Omega_{\rm{b}}h^2$ result, so including BBN in the comparison should not change the conclusion about the (in)consistency. Indeed, now with $N_{\rm{p}}=3$, this high IOI once again exceeds the threshold. The $68\%$-percentile range of the level of physical inconsistency is $\lvb=3.9^{+1.0}_{-1.0}$, again similar to the previous one- and two-parameter cases. 

If we used the commonly adopted $n$-$\sigma$ notation of significance level, then only in the one-parameter case ($H_0$ only) is the $n$-$\sigma$ value equal to the median of $\lvb$ while it is smaller in the two- and three-parameter cases; compare the last row of Table \ref{table:planck-H0}. This example demonstrates that the commonly adopted $n$-$\sigma$ notation of significance level underestimates the level of physical inconsistency when $N_{\rm{p}}>1$ as expected according to the simulations shown in Sec.\,\ref{sec:Bayesian_interpretation_IOI_in_scatters}.

One may argue that $\Omega_{\rm{m}}$ and $\Omega_{\rm{b}}h^2$ are obviously irrelevant in the problem of the Hubble constant tension and should not be counted as an effective parameter in the comparison. But when we compare two constraints in a multiple-parameter model where the two observations have comparable uncertainties for all parameters, we would not know a priori which parameter is relevant and a bias may happen to be in one parameter. Related to this, the common misunderstanding that the ``significance level'' of an inconsistency estimator depends on the number of parameters of the underlying model is rooted in the use of the non-Bayesian interpretation of the estimator and taking it as a presentation of both aspects of physical inconsistency; see the discussion in the problem of quantifying inconsistency in Sec.\,\ref{sec:ambiguity}.

\subsection{Case II: Planck versus DES+BBN}
For the next example, we apply IOI and our new Bayesian interpretation to revisit the (in)consistency between Planck versus the DES year-1 $3\times2$ correlation functions \cite[chains obtained by running the corresponding modules in \textsc{CosmoMC}]{2017-DES-parameter-constraints} in the standard $\Lambda$CDM model. The $3\times2$ correlation functions from DES can put a constraint in the $S_8$-$\Omega_{\rm{m}}$ space, but can only weakly constrain the reduced baryon fraction $\Omega_{\rm{b}}h^2$, the Hubble constant $H_0$ and the scalar spectrum index $n_s$ \cite{2017-DES-parameter-constraints}. In order to show again the effect of increasing the number of parameters that can be constrained by both datasets, we combine DES with the primordial element abundance (BBN) constraint on $\Omega_{\rm{b}}h^2=0.0222\pm0.0005$ \citep{Cooke-etal-2018}, and then investigate the (in)consistency between Planck versus DES+BBN.\footnote{If we used the public MCMC chain for the DES constraint that does not include the BBN prior on $\Omega_{\rm{b}}h^2$, the level of inconsistency would be slightly higher without qualitatively changing our conclusions.} The results are shown in Table \ref{table:planck-DES}. 

The (in)consistency between Planck and DES is usually shown by comparing the two constraints in the $S_8\equiv\sigma_8\times(\Omega_{\rm{m}}/0.3)^{0.5}$ and $\Omega_{\rm{m}}$ plane. Therefore, we first calculate IOI and the range of $\lvb$ in this two-parameter plane, and this is shown in the ($S_8$)+$\Omega_{\rm{m}}$ column of Table \ref{table:planck-DES}. The obtained IOI value is greater than $\rm{IOI}_{\rm{th}}$, which means some physical inconsistency with $\lvb>1$ is likely present. The $68\%$-percentile range of $\lvb$ is $2.4^{+1.0}_{-1.1}$. This is a moderate inconsistency according to the ranking scheme discussed in Sec.\,\ref{sec:guiding-ranking-scheme}. Although based on different inconsistency measures and/or methods of interpretation, our results here agree with Refs.\,\cite{2019-Garcia-Ishak-Fox-Lin,Handley-etal-2019-suspiciousness} in that there is a moderate inconsistency between Planck and DES. 

\begin{table}[tbp]
\centering
\renewcommand{\arraystretch}{1.2}
\begin{tabular}{r|cccc}
\hline\hline
 $S_8$ & +$\Omega_{\rm{m}}$, $N_{\rm{p}}=2$ & +$H_0$, $N_{\rm{p}}=3$ & +$n_s$, $N_{\rm{p}}=4$ & +$\Omega_{\rm{b}}h^2$, $N_{\rm{p}}=5$ \\
 \hline
 $\sqrt{2\rm{IOI}}$ & $2.6>\sqrt{2\rm{IOI}_{\rm{th}}}$ & $2.8>\sqrt{2\rm{IOI}_{\rm{th}}}$ & $3.3>\sqrt{2\rm{IOI}_{\rm{th}}}$ & $3.4>\sqrt{2\rm{IOI}_{\rm{th}}}$ \\
 Range of $\lvb$ & $2.4^{+1.0}_{-1.1}$ & $2.3^{+1.1}_{-1.2}$ & $2.7^{+1.1}_{-1.2}$ & $2.6^{+1.2}_{-1.2}$ \\
 The commonly adopted $n$-$\sigma$ & $2.2$ & $2.0$ & $2.2$ & $2.0$ \\
\hline\hline
\end{tabular}
\caption[Planck vs DES table]{\label{table:planck-DES}Applying IOI and our new Bayesian interpretation to investigate the (in)consistency between Planck versus DES+BBN. The number of parameters increases from $N_{\rm{p}}=2$ ($S_8$, $\Omega_{\rm{m}}$) to $N_{\rm{p}}=5$ ($S_8$, $\Omega_{\rm{m}}$, $H_0$, $n_s$, $\Omega_{\rm{b}}h^2$) as we successively include more parameters in the comparison. We have ${\rm{IOI}}>{\rm{IOI}_{\rm{th}}}$ where we set  $\alpha=0.15$, meaning there likely exists some physical inconsistency.  The level of physical inconsistency $\lvb$ remains in about the same range as inferred from our Bayesian interpretation of IOI, no matter how many more parameters we include in the comparison (as long as doing so does not introduce more physical inconsistency). Comparing the last two rows, we can see that the difference between the commonly adopted $n$-$\sigma$ value and the median of $\lvb$ increases as number of parameters increases. This is another example demonstrating that the non-Baysian $n$-$\sigma$ value underestimates the level of physical inconsistency, and the situation becomes worse as the number of parameters increases. We note that we have combined DES with BBN to set a constraint on $\Omega_{\rm{b}}h^2$.}
\end{table}

If we used the usual significance level, we would have obtained $2.2$-$\sigma$ shown in the last row in Table \ref{table:planck-DES}. This is lower than the median of $\lvb$, indicating that the $n$-$\sigma$ notation tends to underestimate the level of physical inconsistency. 

Next, we successively include more parameters in the comparison. For example, in the +$H_0$ column of Table \ref{table:planck-DES}, we include $H_0$ in the comparison in addition to $S_8$ and $\Omega_{\rm{m}}$, making $N_{\rm{p}}=3$. In the last column, we have in total five parameters in the comparison, including $S_8$, $\Omega_{\rm{m}}$, $H_0$, $n_s$ and $\Omega_{\rm{b}}h^2$. Since $H_0$ is only weakly constrained by DES and the $\Omega_{b}h^2$ constraint from BBN is consistent with that from Planck, we expect that including them in the comparison would not substantially change the inferred level of physical inconsistency (assuming there is no hidden inconsistency in this higher dimension). But including $n_s$ in the comparison somewhat increases the level of $\lvb$, because some hidden inconsistency is caught after doing so \cite{WL2017b}. Nonetheless, all the inferred ranges of $\lvb$ are $\sim2.5\pm1.1$ no matter how many other common parameters we have added to the comparison. On the contrary, the usual significance level gradually drops as we include more parameters in the comparison. This again shows the underestimation of the level of physical inconsistency by the non-Bayesian $n$-$\sigma$ value and it is worse when the number of model parameters increases.  

One may argue that it is not necessary to include $H_0$ in the comparison between Planck and DES+BBN since the latter only weakly constrains it, and those two parameters do not count as extra degrees of freedom in the calculation of the significance level. This type of argument has difficulties to overcome. First of all, it is rather arbitrary to decide whether or not to exclude a common degree of freedom based on how weak one constraint is when compared to the other; see also Ref.\,\cite{2019-Garcia-Ishak-Fox-Lin} for discussion. On the contrary, it is not an issue for IOI and our new Bayesian interpretation. We can see from this example that including the weakly constrained parameters in the comparison actually does not qualitatively affect the result. The reasons are that IOI is stable against noisy eigenmodes (see Ref.\,\cite{2019-Garcia-Ishak-Fox-Lin} for discussion) and that the new interpretation removes the overestimation problem of IOI due to the increase of the parameter number, which is shown in this work. Second, due to degeneracies between parameters, inconsistencies can be hidden, which has been pointed out in \cite{WL2017a,Handley-etal-2019-suspiciousness}. It is therefore important to perform (in)consistency tests in the full common parameter space. This last point also serves as a check on whether it is sufficient to present the (in)consistency between Planck and DES in the $S_8$-$\Omega_{\rm{m}}$ plane only. We can see that the level of physical inconsistency remains in about the same range regardless of whether we choose to calculate $\lvb$ in the $S_8$-$\Omega_{\rm{m}}$ plane or in the full common parameter space. This suggests that the main physical inconsistency is indeed in the $S_8$-$\Omega_{\rm{m}}$ plane.

\subsection{Case III: Planck versus ACT (+WMAP)}
In this last case study, we apply IOI and our Bayesian interpretation to revisit the (in)consistency between the constraints from Planck and the CMB temperature and polarization anisotropy from the Atacama Cosmology Telescope \cite[chains obtained from \href{https://lambda.gsfc.nasa.gov}{https://lambda.gsfc.nasa.gov}]{2020-ACT-parameters} in the standard cosmological model in the six-parameter space, i.e., $\Omega_{b}h^2$, $\Omega_{c}h^2$, $\theta$, $A_s$, $n_s$ and $\tau$.

\begin{table}[]
    \centering
    \renewcommand{\arraystretch}{1.2}
    \begin{tabular}{r|ccc}
    \hline\hline
           &  $\sqrt{2\rm{IOI}}$  & $\lvb$ & Commonly adopted $n$-$\sigma$  \\
        \hline
        Planck vs. ACT & $4.1>\sqrt{2\rm{IOI}_{\rm{th}}}$ &  $3.3^{+1.1}_{-1.2}$ & $2.6$  \\
        Planck vs. ACT+WMAP & {\color{blue}$2.1<\sqrt{2\rm{IOI}_{\rm{th}}}$}  & $1.1^{+1.1}_{-0.8}$ & $0.5$\\
    \hline\hline
    \end{tabular}
    \caption{Applying IOI and our new Bayesian interpretation to revisit the (in)consistency between Planck versus ACT in the standard $\Lambda$CDM cosmological model. For Planck versus ACT alone, there exists a moderate physical inconsistency. However, including WMAP in the ACT analysis restores its consistency with Planck as the corresponding IOI value (marked blue) does not exceed the threshold. In either case, using the commonly adopted $n$-$\sigma$ notation significantly underestimates the level of physical inconsistency.  }
    \label{table:Planck-ACT}
\end{table}

We first compare Planck versus the ACT-only constraint. The results are shown in the Planck vs. ACT row in Table \ref{table:Planck-ACT}. We can see that IOI$>$IOI$_{\rm{th}}$ and the level of physical inconsistency likely exceeds the threshold. In fact, the probability for $\lvb>1$ is $96.5\%$. The $68\%$-percentile range of the level of physical inconsistency is $3.3^{+1.1}_{-1.2}$. This is a moderate-to-strong inconsistency according to the ranking scheme discussed in Sec.\,\ref{sec:guiding-ranking-scheme}. We also reproduce the result reported in Ref.\,\cite{2020-Handley-Lemos} which uses the commonly adopted significance level, $2.6$-$\sigma$, for this comparison\footnote{We note that to reproduce this $2.6$-$\sigma$ significance level, we have used our IOI and Eqs.\,\eqref{eq-convert-IOI-PTE} and \eqref{eq-significance-P-IOI} instead of the suspiciousness statistics used in Ref.\,\cite{2020-Handley-Lemos}. We can do so because in the Gaussian limit IOI and suspiciousness differ by only a constant offset \cite{Handley-etal-2019-suspiciousness}.}. While we agree with Ref.\,\cite{2020-Handley-Lemos} that some certain physical inconsistency exists between Planck and ACT, we emphasize that this $2.6$-$\sigma$ should \emph{not} be interpreted as the level of physical inconsistency since it is significantly lower than the median of $\lvb$. This once again demonstrates the underestimation of the level of physical inconsistency by the commonly adopted $n$-$\sigma$ notation of significance level as expected according to the simulations shown in Sec.\,\ref{sec:Bayesian_interpretation_IOI_in_scatters}.

Next, we compare Planck versus ACT+WMAP. As pointed out in Ref.\,\cite{2020-ACT-parameters}, ACT is unable to probe the large-scale (or small angular multipole) CMB anisotropy and requires a prior on the amplitude of the large-scale CMB (e.g., temperature) spectrum. Thus, Ref.\,\cite{2020-ACT-parameters} also produced the joint constraint using ACT and WMAP. The results of (in)consistency between Planck versus ACT+WMAP are shown in the last row of Table \ref{table:Planck-ACT}. We can see that now IOI does not exceed the threshold value meaning that there is no strong evidence for the existence of any substantial physical inconsistency between Planck versus ACT+WMAP. This is consistent with Ref.\,\cite{2020-ACT-parameters} where including the prior from WMAP on the large-scale CMB spectrum amplitude restores the consistency between Planck and ACT. To comment on this result, we note that on the one hand it is unsafe to have some unreasonable prior \,\cite{2008-Efstathiou-limit-of-Bayesian-evidence}. On the other hand, this may again point to the discrepancy between the large scales and the small scales in the framework of the standard $\Lambda$CDM cosmological model \cite{Addison-etal-2016}. Although it was claimed that this discrepancy has been addressed \cite{Planck-XI-2016}, it is interesting to thoroughly reanalyze this issue using IOI and our new Bayesian interpretation. While this is beyond the scope of this work, we will leave this to a dedicated work in the future.

\section{An ambiguity in the quantification of inconsistency in multi-parameter cases and the resolution in our Bayesian interpretation}\label{sec:ambiguity}
We have discussed a Bayesian interpretation of the IOI values, which is based on the conditional probability distribution of the level of physical inconsistency given the obtained value of IOI. The new interpretation is summarized into two questions described in Sec.\,\ref{sec:Bayesian_interpretation_IOI_in_scatters} and is thus twofold. The first question is to evaluate the probability for the level of physical inconsistency to exceed the threshold (we call it ``the threshold-exceeding probability'' in this section). The commonly adopted PTE and the associated $n$-$\sigma$ notation of confidence level has a similar objective, but they are based on the probability distribution of an inconsistency estimator in the absence of any physical inconsistency and thus (at best) only indirectly address the desired task (i.e., our first question). In either case, this evaluation is only to see whether the level of the physical inconsistency has reached some threshold. Surely, a probability of $99\%$ is larger than a probability of $90\%$, but once there is likely to be a non-negligible physical inconsistency, we should then ask what the actual level of physical inconsistency is, which directly relates to the extent to which the physical inconsistency biases/affects the constraints. This leads to our second question, which is to estimate the range of the actual level of physical inconsistency. It is the range of the level of physical inconsistency that is more important. 

In one-parameter cases both aspects above increase with the value of IOI (or some other similar inconsistency estimators). Thus, when the interpretation of these two aspects are suitably adjusted, we can ``conveniently'' specify the two aspects using one quantity (like the usual n-$\sigma$ notation). Such a ``convenience'' no longer holds in multiple-parameter cases. This is because the two aspects behave very differently as $N_{\rm{p}}$ increases. For a given IOI, while the threshold-exceeding probability drops with $N_{\rm{p}}$, the range of the level of physical inconsistency only weakly depends on $N_{\rm{p}}$, especially when IOI is large. 

The lack of understanding of the distinction between the two aspects leads to an ambiguity in the quantification of inconsistency in multiple-parameter cases. Regarding this, there have been two different and conflicting interpretations of inconsistency estimators in the literature, each of which emphasizes one aspect of the quantification of inconsistency but overlooks the other. On the one hand, measures of inconsistency in the literature often use the common ``significance level'', which \emph{only indirectly} quantifies how likely it is for a physical inconsistency to be present but gives a misleading understanding of the actual level of physical inconsistency.\footnote{There have been techniques that remove irrelevant parameter degrees of freedom and alleviate (but not eliminate) the problem of underestimation, e.g., see Ref.\,\cite{2018-Raveri-Hu,Handley-etal-2019-suspiciousness}. Our new Bayesian framework is clearly more straightforward.} On the other hand, the original interpretation of IOI that uses the Jeffrey's scales tries to describe the level of physical inconsistency \cite{WL2017a}, but overlooks the drop (with the parameter number) of threshold-exceeding probability. Our new Bayesian interpretation unifies the two different aspects because they (and other information) are all included in the conditional probability distribution introduced (i.e. $P(\lvb|\sqrt{2\rm{IOI}})$). 

One may suspect that the range of the level of physical inconsistency being insensitive to $N_{\rm{p}}$ contradicts the drop of the threshold-exceeding probability with $N_{\rm{p}}$. There is actually no contradiction here. Let us take a large-IOI limit for discussion. At such a limit, the median of $\lvb$ given an obtained IOI can be well-approximated by $\sqrt{2\rm{IOI}}$ (the term $N_{\rm{p}}-1$ can be ignored), and the most probable ($68\%$-percentile) range of $\lvb$ can be well-approximated by $\sqrt{2\rm{IOI}}\pm1$. The main effect from $N_{\rm{p}}$ on the distribution of $\lvb$ is not in the most probable range of $\lvb$, but in the tail of the distribution in the range of small $\lvb$ values. As a consequence, the probability 1-$P(\lvb>1|\sqrt{2\rm{IOI}})$ (which determines whether $P(\lvb>1|\sqrt{2\rm{IOI}})$ is $95$\% or $99$\%) has a much larger dependence on $N_{\rm{p}}$ as it is sensitive to the distribution of $\lvb$ at the low value range. But note that while we emphasize that the second question regarding the most probable range of $\lvb$ is more important, it does not mean the first question is unimportant. This is because the probability for the level of physical inconsistency to reach the threshold needs to be high enough for it to be further considered.

\section{Conclusion and outlook}\label{sec:IOI-conclusion-outlook}
We propose a Bayesian interpretation of inconsistency measures based on the conditional probability distribution of the level of physical inconsistency given an obtained value of the measure. We use the index of inconsistency as a case study. The statistical properties of the level of physical inconsistency are then presented by two summary statistics: (1) the probability of the level of physical inconsistency to reach the threshold, and (2) the $68\%$  (or $95\%$) percentile range of the level of physical inconsistency. While the first only serves as a threshold criterion, the second is more important and directly relates to the extent to which the physical inconsistency biases the cosmological constraints. 

A key concept in this work is that cosmological inconsistencies can be related to physical sources such as model problems or unaccounted for systematic effects, or due to scatter in the data. This new framework allows one to properly extract information about the physical inconsistency in the presence of data scatter.   

We show with real data that the commonly adopted $n$-$\sigma$ notation for the significance tends to underestimate the actual level of physical inconsistency. The underlying reason is that the two summary statistics above have different behaviors as the number of model parameters increases. On the one hand, the probability of the level of  physical inconsistency to reach the threshold is sensitive to the tail of the distribution and drops as the number of model parameters increases. On the other hand, the most probable range of the level of physical inconsistency remains rather insensitive to the number of model parameters, especially when the physical inconsistency is large. These two different behaviors cause a common ambiguity in the quantification of inconsistency. The commonly adopted $n$-$\sigma$ notation only \emph{indirectly} reflects the first aspect but overlooks the second. The proposed Bayesian interpretation unifies these two aspects and resolves such an ambiguity. To our knowledge, this Bayesian interpretation of inconsistency measures is the first work of this type.  

Although we use IOI as a case study, this framework can be generalized to other inconsistency measures. The general steps to apply this Bayesian interpretation to other measures are: (1) Find the proper inconsistency measures of physical inconsistency in a general situation in the absence of data scatter; (2) Use Bayes' theorem to calculate the conditional probability distribution of the physical inconsistency to take into account data scatter.

Looking ahead, a plethora of precise data will be delivered in the near future, such as galaxy surveys from  LSST, DESI, Euclid, WFIRST and SKA, and CMB missions such as, for example, CORE and CMB Stage-IV\footnote{\label{ft:surveys}See   \href{https://www.desi.lbl.gov/}{https://www.desi.lbl.gov/} for DESI,
\href{http://www.lsst.org/lsst}{http://www.lsst.org/lsst} for LSST, 
\href{http://sci.esa.int/euclid/}{http://sci.esa.int/euclid} for Euclid, \href{https://wfirst.gsfc.nasa.gov}{https://wfirst.gsfc.nasa.gov} for WFIRST, \href{https://www.skatelescope.org}{https://www.skatelescope.org} for SKA, \href{https://www.cosmos.esa.int/}{https://www.cosmos.esa.int/} for CORE, and \href{https://cmb-s4.org}{https://cmb-s4.org} for Stage-IV.}. Additionally, gravitational waves have opened a new window providing even more fundamentally different methods to probe our universe. Comparing results from different observations and examining their (in)consistency is thus quickly becoming very essential. Our work provides a new framework to assess physical inconsistencies in the presence of data scatter which is an important step in the study of cosmological inconsistencies.

\acknowledgments
NOTE: An earlier version-2 of this paper was resubmitted to JCAP on May 13th 2020 where we proposed our new framework of the Bayesian interpretation of inconsistency measures along with the concept of separating physical inconsistency and data scatter, but was not posted to the arXiv server. While working on shortening version-2 of the paper for resubmission to JCAP, similar concepts to physical inconsistency and separating between physical inconsistency and data scatter appeared on the arXiv in \cite{2020-Lemos-etal} on December 17th 2020. 

We thank Orion Ning for proofreading the manuscript, and Logan Fox and Cristhian Garcia-Quintero for useful comments. We also thank an anonymous referee for useful comments. MI acknowledges that this material is based upon work supported in part by the U.S. Department of Energy, Office of Science, under Award Numbers DE-SC0019206 and DE-SC0021277.

\bibliographystyle{JHEP}
\bibliography{IOIremarks}

\end{document}